\documentclass[12 pt]{article}

\pdfoutput=1

\usepackage{fullpage}
\usepackage{amscd}
\usepackage{amsmath}
\usepackage{amssymb}
\usepackage{amsthm}
\usepackage{graphicx}
\usepackage{latexsym}
\usepackage{verbatim}
\input epsf.tex

\newtheorem{theorem}{Theorem}[section]
\newtheorem{lemma}[theorem]{Lemma}
\newtheorem{corollary}[theorem]{Corollary}

\newtheorem{proposition}[theorem]{Proposition}

\theoremstyle{definition}
\newtheorem{assumption}[theorem]{Assumption}

\theoremstyle{definition}
\newtheorem{algorithm}[theorem]{Algorithm}

\theoremstyle{remark}
\newtheorem{example}[theorem]{Example}

\theoremstyle{definition} 
\newtheorem{definition}[theorem]{Definition}

\newcounter{tempthm}
\newcounter{tempsec}

\newcommand{\savecounter}[1]{\newcounter{thmcounter#1}
\setcounter{thmcounter#1}{\value{theorem}}
\newcounter{seccounter#1}
\setcounter{seccounter#1}{\value{section}}}

\newcommand{\usesavedcounter}[1]{\setcounter{tempthm}{\value{theorem}}
\setcounter{theorem}{\value{thmcounter#1}}
\setcounter{tempsec}{\value{section}}
\setcounter{section}{\value{seccounter#1}}}

\newcommand{\restorecounter}{\setcounter{theorem}{\value{tempthm}}
\setcounter{section}{\value{tempsec}}}

\newcommand{\Rm}{\mathbb{R}}

\newcommand{\shortv}[1]{}

\DeclareMathOperator{\kernel}{ker}

\DeclareMathOperator{\linspan}{span}
\DeclareMathOperator{\rank}{rank}
\DeclareMathOperator{\conv}{conv}
\DeclareMathOperator{\aff}{aff}

\DeclareMathOperator{\supp}{supp}

\title{\LARGE \bf
Separable and Low-Rank Continuous Games
}

\author{Noah D. Stein, Asuman Ozdaglar, and Pablo A. Parrilo
\thanks{Department of Electrical Engineering,
        Massachusetts Institute of Technology: Cambridge, MA 02139.
        {\tt\small nstein@mit.edu}, {\tt\small asuman@mit.edu}, and {\tt\small parrilo@mit.edu}.}
\thanks{This research was funded in part by National Science Foundation grants $\text{DMI-}0545910$ and ECCS-$0621922$ and AFOSR MURI subaward $2003\text{-}07688\text{-}1$.}
      }

\begin{document}

\markright{LIDS Technical Report $2760$}

\maketitle

\thispagestyle{headings}

\pagestyle{plain}

\begin{abstract}

In this paper, we study nonzero-sum separable games, which are
continuous games whose payoffs take a sum-of-products form. Included
in this subclass are all finite games and polynomial games. We
investigate the structure of equilibria in separable games. We show
that these games admit finitely supported Nash equilibria. Motivated
by the bounds on the supports of mixed equilibria in two-player
finite games in terms of the ranks of the payoff matrices, we define
the notion of the rank of an $n$-player continuous game and use this
to provide bounds on the cardinality of the support of
equilibrium strategies. We present a general characterization
theorem that states that a continuous game has finite rank if and only if
it is separable. Using our rank results, we present an efficient
algorithm for computing approximate equilibria of two-player
separable games with fixed strategy spaces in time polynomial in the rank of the game.

\end{abstract}

\section{Introduction}

The structure and computation of equilibria in games with infinite strategy spaces have long been recognized as complex.  Even seemingly ``good'' games may possess only ``bad'' equilibria; Gross has constructed an example of a zero-sum game with rational utility functions whose unique Nash equilibrium is for each player to play the Cantor measure \cite{g:rpccd,karlin:tig}.  To avoid such pathologies,  Dresher, Karlin, and Shapley introduced the class of zero-sum separable games \cite{dks:pg,dk:scgfp,ks:gms,karlin:tig}.  These are games in which each player's payoff can be written as a sum of products of functions in each player's strategy separately (e.g.\ as polynomials), and it is known that every separable game admits a mixed strategy Nash equilibrium that is finitely supported, i.e.\ all players randomize over a finite number of pure strategies.  Parrilo has shown that Nash equilibria of zero-sum games with polynomial utility functions can be computed efficiently \cite{pp:polygames}.

In this paper, we study nonzero-sum separable games.  We show that even in the nonzero-sum case, separable games have the property that an equilibrium exists in finitely supported mixed strategies.  We then characterize the structure of these games and their Nash equilibria, and also propose methods for computing (exact and approximate) equilibria.

Our first major contribution is to define and develop the concept of the \textit{rank of a continuous game}, which we use to construct bounds on the number of strategies played in equilibrium.  In two-player finite games, Lipton et al.\ have recently shown that the ranks of the payoff matrices provide such a bound \cite{lmm:plgss}. Our new definition of rank generalizes this one to allow for an arbitrary finite number of players (a problem explicitly left open in \cite{lmm:plgss}) and infinite strategy spaces.  We define the rank of a continuous game by introducing an equivalence relation between mixed strategies called \textit{almost payoff equivalence}.  The rank is the dimension of the mixed strategy space modulo this equivalence relation. Loosely speaking, low-rank continuous games are those where the variation in each player's payoff depends only on a low-dimensional projection of the mixed strategies of the players.  For example in games with polynomial payoffs, the rank depends on the dimension of a projection of the moment space.

We also show that a continuous game has finite rank if and only if it is separable. This means that little can be said about the structure of equilibria in non-separable continuous games and highlights the importance of separable games.  We provide simple techniques for evaluating the rank of separable games, which we specialize to games with polynomial payoffs and finite games.

Our second set of results concern efficient computation of mixed
strategy equilibria in separable games. For $n$-player games, we
provide a nonlinear optimization formulation, and show that the
optimal solutions of this problem correspond to the (generalized)
moments of exact Nash equilibria. This formulation generalizes the
optimization formulation of Nash equilibria of finite games presented by Ba\c{s}ar and Olsder \cite{bo:dngt}.

While the nonlinear optimization formulation for the computation of
equilibria is tractable for certain classes of separable games such
as zero-sum polynomial games, the potential nonconvexity of the
optimization problem makes it impractical in other cases. We
therefore supplement our computational results for exact mixed
strategy equilibria with new methods for computing approximate
mixed strategy equilibria. Using our rank results described above,
we can link the computation of mixed strategy equilibria to the
appropriate discretization of the strategy space into finite
actions. For two-person separable games, this yields an
efficient algorithm for computing  approximate mixed strategy
equilibria in time polynomial in the rank of the game. This
algorithm searches for finitely supported approximate equilibria by
enumerating all possible supports, and relies on the fact that the
search can be limited to supports of size bounded by the rank of the game.
For two-players, the set of equilibria for a given support can be
described by linear equations and inequalities \footnote{With
more than two players this description is no longer linear (or
convex), so this algorithm does not generalize to $n$-player games with
$n>2$.}.

Our work is related to a number of literatures. There has been
considerable work on the computation of equilibria in finite games.  Lemke and Howson give a path-following algorithm for two-player finite games which can be viewed as the simplex method for linear programming operating with a different pivoting rule \cite{lh:epbg}.  To find equilibria of games with more players, Scarf constructs a simplicial subdivision algorithm which also works for more general fixed point problems \cite{s:afpcm}.  These methods rely on the polyhedral structure of the mixed strategy spaces of finite games, therefore they seem unlikely to generalize to continuous/separable games.  For a survey of algorithms which compute equilibria of finite games, see \cite{mm:ce}.

Another growing literature has been investigating the complexity of
computing mixed strategy Nash equilibria of finite games.  Daskalakis, Goldberg, and Papadimitriou settle this question for finite normal form games with four or more players, showing that the problem of computing a single Nash equilibrium is PPAD-complete \cite{dgp:ccne}.  In essence this means that it is computationally equivalent to a number of other fixed point problems which are believed to be computationally difficult.  These problems share the feature that a solution can be proven to exist, but the known proofs of existence are inefficient; for more about the complexity class PPAD, see \cite{p:cpa}.  Daskalakis and Papadimitriou later improve this result by proving PPAD-completeness in the case of three players \cite{dp:3nash}.  Chen and Deng also prove this independently \cite{cd:3nash} and complete this line of work by proving PPAD-completeness for two players \cite{cd:nashcomp}.  In this
literature, there has been no work on continuous games.

Our work is most closely related to the work of Lipton et al.\
\cite{lmm:plgss}, who consider two-player finite games and
provide bounds on the support of equilibrium strategies using the
ranks of the payoff matrices of the players. Since finite games are a
special case of separable games, our results on the cardinality of
the support of equilibrium strategies generalize theirs by allowing
for an arbitrary finite number of players as well as infinite
strategy sets and a broader class of payoff functions.

Lipton et al.\ also investigate computing approximate equilibria in two-player
finite games and present the first algorithm for computing
approximate equilibria which is quasi-polynomial in the number of
strategies \cite{lmm:plgss}. The rank of a separable game measures the complexity of the payoffs, and in the case of a finite game it is bounded by the number of strategies. Therefore in finite games the complexity of the payoffs and the complexity of the strategy spaces do not vary independently; the effects of these parameters on the running time of algorithms cannot be distinguished.  However, for games with infinite strategy sets the rank can be varied arbitrarily while the strategy set is held fixed.  This allows us to construct an algorithm for computing approximate equilibria in two-player separable games with fixed (infinite) strategy spaces and show that it has a polynomial dependence on the rank.  Since we assume the strategy space is fixed and consider general separable games rather than only finite games, our algorithm is not directly comparable to that of Lipton et al.\ \cite{lmm:plgss}.  Nonetheless it is interesting on its own since no algorithm for computing approximate equilibria of finite games is known which has polynomial dependence on the complexity of the game.

Our work is also related to that of Kannan and Theobald, who study a different notion of rank in two-player finite games \cite{kt:gfr}.  They take an algorithmic perspective and view zero-sum games as the simplest type of games.  To generalize these, they propose a hierarchy of classes of two-player finite games in which the rank of the sum $R+C$ of the payoff matrices is bounded by a constant $k$; the case $k=0$ corresponds to zero-sum games.  For fixed $k$, they show that approximate Nash equilibria of two-player finite games can be computed in time polynomial in the description length for the game.  This algorithm relies on an approximation result for quadratic programming due to Vavasis \cite{v:aaiqp} which depends on the polyhedral structure of the problem.  It is conceivable that this technique may apply to polynomial games if the approximation technique can be extended to this more general algebraic setting, but we do not do so here.

The rest of this paper is organized as follows. In Section
\ref{sec:basictheory} we define separable games and prove some of their basic properties.  Then in Section \ref{sec:rank} we define the rank of a
continuous game and use this definition to bound the cardinality of the support of Nash
equilibria. We present a characterization theorem for separable
games which in particular shows that within the class of continuous
games, the low-rank results in this paper cannot be extended past
separable games. We provide a simple formula for computing the rank
of arbitrary separable games, which we specialize for finite and
polynomial games. In Section \ref{sec:comp} we discuss computation
of Nash equilibria and approximate equilibria. We close with
conclusions and directions for future work.

\section{Basic Theory of Separable Games}

\label{sec:basictheory}

We first introduce the notational conventions and the basic
terminology used throughout the paper.  Subscripts refer to
players, while superscripts are reserved for other indices, rather
than exponents. If $S_j$ are sets for $j=1,\ldots,n$ then $S =
\Pi_{j=1}^n S_j$ and $S_{-i} = \Pi_{j\neq i} S_j$.  The $n$-tuple
$s$ and the $(n-1)$-tuple $s_{-i}$ are formed from the points $s_j$
similarly. Given a subset $S$ of a vector space, we use the notation $\linspan S$, $\aff
S$, $\conv S$, and $\overline{S}$ to denote the span, affine hull,
convex hull, and closure of the set $S$, respectively. We denote the transpose of a
matrix $M$ by $M'$.

\begin{definition}
An $n$-player \textbf{continuous game} consists of a \textbf{pure strategy} space $C_i$ which is a nonempty compact metric space and a continuous \textbf{utility} or \textbf{payoff function}
$u_i: C\rightarrow\Rm$ for each player $i = 1,\ldots, n$.  Throughout,
$\Delta_i$ will denote the space of Borel probability measures
$\sigma_i$ over $C_i$, referred to as \textbf{mixed strategies}, and
the $u_i$ will be extended from $C$ to $\Delta$ by expectation,
defining
\begin{equation*}
  u_i(\sigma) = \int_C u_i(s) d\sigma,
\end{equation*}
where the $n$-tuple $\sigma = (\sigma_1,\ldots,\sigma_n)$ is identified with the product measure $\sigma_1\times\cdots\times\sigma_n$.
\end{definition}

\begin{definition}
An $n$-player \textbf{separable game} is an $n$-player continuous
game with utility functions $u_i: C\rightarrow\Rm$ taking the form
\begin{equation}
\label{eq:sepform}
    u_i(s) = \sum_{j_1=1}^{m_1} \cdots \sum_{j_n=1}^{m_n} a_i^{j_1\cdots j_n}
    f_1^{j_1}(s_1)\cdots f_n^{j_n}(s_n),
\end{equation}
where $a_i^{j_1\cdots j_n}\in\Rm$ and the $f_i^j: C_i\rightarrow\Rm$
are continuous.  Two important special cases are the \textbf{finite
games} in which the $C_i$ are finite and the $u_i$ are arbitrary and
the \textbf{polynomial games} in which the $C_i$ are compact intervals in
$\Rm$ and the $u_i$ are polynomials in the $s_j$.
\end{definition}

When it is convenient to do so, and always for polynomial games, we
will begin the summations in \eqref{eq:sepform} at $j_i = 0$.  For
polynomial games we can then use the convention that $f_i^j(s_i) =
s_i^j$, where the superscript on the right hand side denotes an
exponent rather than an index.

\savecounter{ex1}
\begin{example}
\label{example:ex1}
Consider a two player game with $C_1 = C_2 = [-1,1]\subset\Rm$.
Letting $x$ and $y$ denote the pure strategies of players $1$ and
$2$, respectively, we define the utility functions
\begin{equation}
\label{eq:expayoffs}
\begin{split}
    u_1(x,y) &= 2xy + 3y^3 - 2x^3 - x - 3x^2 y^2\text{\ \ and} \\
    u_2(x,y) &= 2x^2 y^2 - 4y^3 - x^2 + 4y + x^2 y.
\end{split}
\end{equation}
This is a polynomial game, and we will return to it periodically to
apply the results presented.  In particular, we will show using classical techniques that this game must have a Nash equilibrium in which each player randomizes over a set of cardinality at most $5$.  We will then apply our new rank results (see Theorem \ref{thm:seprank}) to reduce this bound to $2$ for the first player and $4$ for the second player.
\end{example}

Let $V_i$ denote the space of all finite-valued signed measures
(henceforth simply called measures) on $C_i$, which can be
identified with the dual of the Banach space $C(C_i)$ of all
continuous real-valued functions on $C_i$ endowed with the sup norm.
Throughout, we will use the \textbf{weak*} topology on $V_i$, which
is the weakest topology such that whenever $f: C_i\rightarrow\Rm$ is a
continuous function, $\sigma\mapsto\int f d\sigma$ is a continuous linear
functional on $V_i$.

We can extend the utility functions of a continuous game to all of
$V$ in the same way they are extended from $C$ to $\Delta$, yielding
a multilinear functional on $V$.  For a fixed separable game we can
extend the $f_i^j$ from $C_i$ to $V_i$ similarly, yielding the
so-called \textbf{generalized moment} functionals, so
\eqref{eq:sepform} holds with $s$ replaced by $\sigma\in V$.  In polynomial games $f_i^j(s_i) = s_i^j$ so the generalized
moment functionals are just the classical moment functionals.  We
will abuse notation and identify the elements of $C_i$ with the
atomic measures in $\Delta_i$, so $C_i\subseteq\Delta_i\subset V_i$.
Note that $\conv C_i$ and $\linspan C_i$ are the set of all finitely
supported probability measures and the set of all finitely supported
finite signed measures, respectively.  The following are standard
results (see \cite{p:pmms} and \cite{r:fa}).
\begin{proposition}
\label{prop:standardtop}
\mbox{}
\begin{itemize}
    \item[(a)] The sets $C_i$ and $\Delta_i$ are weak* compact.
    \item[(b)] The weak* closures of $\conv C_i$ and $\linspan C_i$ are $\Delta_i$ and $V_i$, respectively.
    \item[(c)] The weak* topology makes $V_i$ a Hausdorff topological vector space.
\end{itemize}
\end{proposition}

We next define three notions of equivalence between two measures,
which allow us to exhibit simplifications in the structure of Nash
equilibria in separable games.

\begin{definition}
\label{def:equivrels}
Two measures $\sigma_i,\tau_i\in V_i$ are
\begin{itemize}
    \item \textbf{moment equivalent} if $f_i^j(\sigma_i) = f_i^j(\tau_i)$ for all $j$ (representation-dependent and only defined for separable games).
    \item \textbf{payoff equivalent} if $u_j(\sigma_i,s_{-i}) = u_j(\tau_i,s_{-i})$ for all $j$ and all $s_{-i}\in C_{-i}$.
    \item \textbf{almost payoff equivalent} if $u_j(\sigma_i,s_{-i}) = u_j(\tau_i,s_{-i})$ for all $j\neq i$ and all $s_{-i}\in C_{-i}$.
\end{itemize}
\end{definition}

Note that in separable games moment equivalence implies payoff
equivalence and in all continuous games payoff equivalence implies
almost payoff equivalence.  Since the $f_i^j$ and $u_j$ are linear
and multilinear functionals on $V_i$ and $V$, respectively, these
equivalence relations can be expressed in terms of (potentially
infinitely many) linear constraints on $\sigma_i - \tau_i$.

\begin{definition}Let $0$ denote the zero measure in $V_i$ and define
\begin{itemize}
\item $W_i = \left\{\text{measures moment equivalent to }0\right\}$,
\item $X_i = \left\{\text{measures payoff equivalent to }0\right\}$,
\item $Y_i = \left\{\text{measures almost payoff equivalent to }0\right\}$.
\end{itemize}
\end{definition}

Then $W_i\subseteq X_i\subseteq Y_i$, and $\sigma_i - \tau_i\in X_i$ if and only if $\sigma_i$ is payoff equivalent to $\tau_i$, etc.  Furthermore, the subspaces $X_i$ and $Y_i$ are representation-independent and well-defined for any continuous game, separable or not.  Note that these subspaces are given by the intersection of the kernels of (potentially infinitely many) continuous linear functionals, hence they are closed.

We will analyze separable games by considering the quotients of
$V_i$ by these subspaces, i.e. $V_i$ mod these three equivalence
relations.  To avoid defining excessively many symbols let
$\Delta_i/W_i$ denote the image of $\Delta_i$ in $V_i/W_i$ and so
forth.  For a concrete example of these sets see the discussion at end of this section, and in particular Figure \ref{fig:momentspace}.

The following theorem presents a fundamental result about separable
games.  It shows that regardless of the choices of the other
players, each player is free to restrict his choice of strategies to
a particularly simple class of mixed strategies, namely those which
only assign positive probability to a finite number of pure
strategies.  Furthermore, a bound on the number of strategies needed
can be easily computed in terms of the structure of the game.  This
theorem can be proven by a separating hyperplane argument as applied
to zero-sum separable games by Karlin \cite{karlin:tig}, but here we
give a new topological argument.

\begin{theorem}
\label{thm:seppayoffequiv} In a separable game every mixed strategy
$\sigma_i$ is moment equivalent to a finitely-supported mixed
strategy $\tau_i$ with $|\supp(\tau_i)|\leq m_i + 1$. Moreover, if
$\sigma_i$ is countably-supported $\tau_i$ can be chosen with
$\supp(\tau_i)\subset\supp(\sigma_i)$.
\end{theorem}

\begin{proof}
Note that the map
\begin{equation}
\label{eq:momentfunction}
f_i: \sigma_i\mapsto \left(f_i^1(\sigma_i),\ldots,f_i^{m_i}(\sigma_i)\right)
\end{equation}
is linear and continuous.  Therefore
\begin{equation*}
    f_i(\Delta_i) = f_i\left(\overline{\conv C_i}\right)\subseteq \overline{f_i(\conv C_i)} = \overline{\conv f_i(C_i)} = \conv f_i(C_i) = f_i(\conv C_i)\subseteq f_i(\Delta_i).
\end{equation*}
The first three steps follow from Proposition
\ref{prop:standardtop}, continuity of $f_i$, and linearity of $f_i$,
respectively.  The next equality holds because $\conv f_i(C_i)$ is
compact, being the convex hull of a compact subset of a
finite-dimensional space.  The final two steps follow from the
linearity of $f_i$ and the containment $\conv C_i\subseteq\Delta_i$.
Hence, we have
\[f_i(\Delta_i) = \conv f_i(C_i) = f_i(\conv C_i).\]
This shows that any mixed strategy is moment equivalent to a
finitely-supported mixed strategy. Applying Carath\'{e}odory's
theorem \cite{bno:convex} to the set $\conv
f_i(C_i)\subset\Rm^{m_i}$ yields the uniform bound.  Since a
countable convex combination of points in a bounded subset of
$\Rm^{m_i}$ can always be written as a finite convex combination of
at most $m_i+1$ of those points, the final claim follows.
\end{proof}

For the rest of the paper we will study the Nash equilibria of
(nonzero-sum) separable games, which are defined for arbitrary continuous games
as follows.

\begin{definition} A mixed strategy profile $\sigma$ is a
\textbf{Nash equilibrium} if $u_i(\tau_i,\sigma_{-i})\leq u_i(\sigma)$ for all $i$ and all $\tau_i\in\Delta_i$.
\end{definition}

Combining Theorem \ref{thm:seppayoffequiv} with Glicksberg's result
\cite{glicksberg:cg} that every continuous game has a Nash
equilibrium yields the following:

\begin{corollary}
\label{cor:sepeq} Every separable game has a Nash equilibrium in
which player $i$ mixes among at most $m_i + 1$ pure strategies.
\end{corollary}

\usesavedcounter{ex1}
\begin{example}[cont'd] Apply the standard definition of the $f_i^j$ to the polynomial game with payoffs given in \eqref{eq:expayoffs}.  The set of moments $\Delta_i/W_i \cong f_i(\Delta_i)$ as described in Theorem \ref{thm:seppayoffequiv} is shown in Figure \ref{fig:momentspace} with the zeroth moment omitted (it is identically unity).  In this case the set of moments is the same for both players.  The space $V_i/W_i = f_i(V_i)$ is a four-dimensional real vector space, and Figure \ref{fig:momentspace} shows a subset of the projection of $V_i/W_i$ onto its final three coordinates.  The set $C_i/W_i = f_i(C_i)$ of moments due to pure strategies is the curve traced out by the vector $(1,x,x^2,x^3)$ as $x$ varies from $-1$ to $1$.  Since the first coordinate is omitted in the figure, this can be seen as the sharp edge which goes from $(-1,1,-1)$ through $(0,0,0)$ to $(1,1,1)$.  As shown in the proof of Theorem \ref{thm:seppayoffequiv}, $f_i(\Delta_i)$ is exactly the convex hull of this curve.  

\begin{figure}
    \centering
        \includegraphics[width=0.70\textwidth]{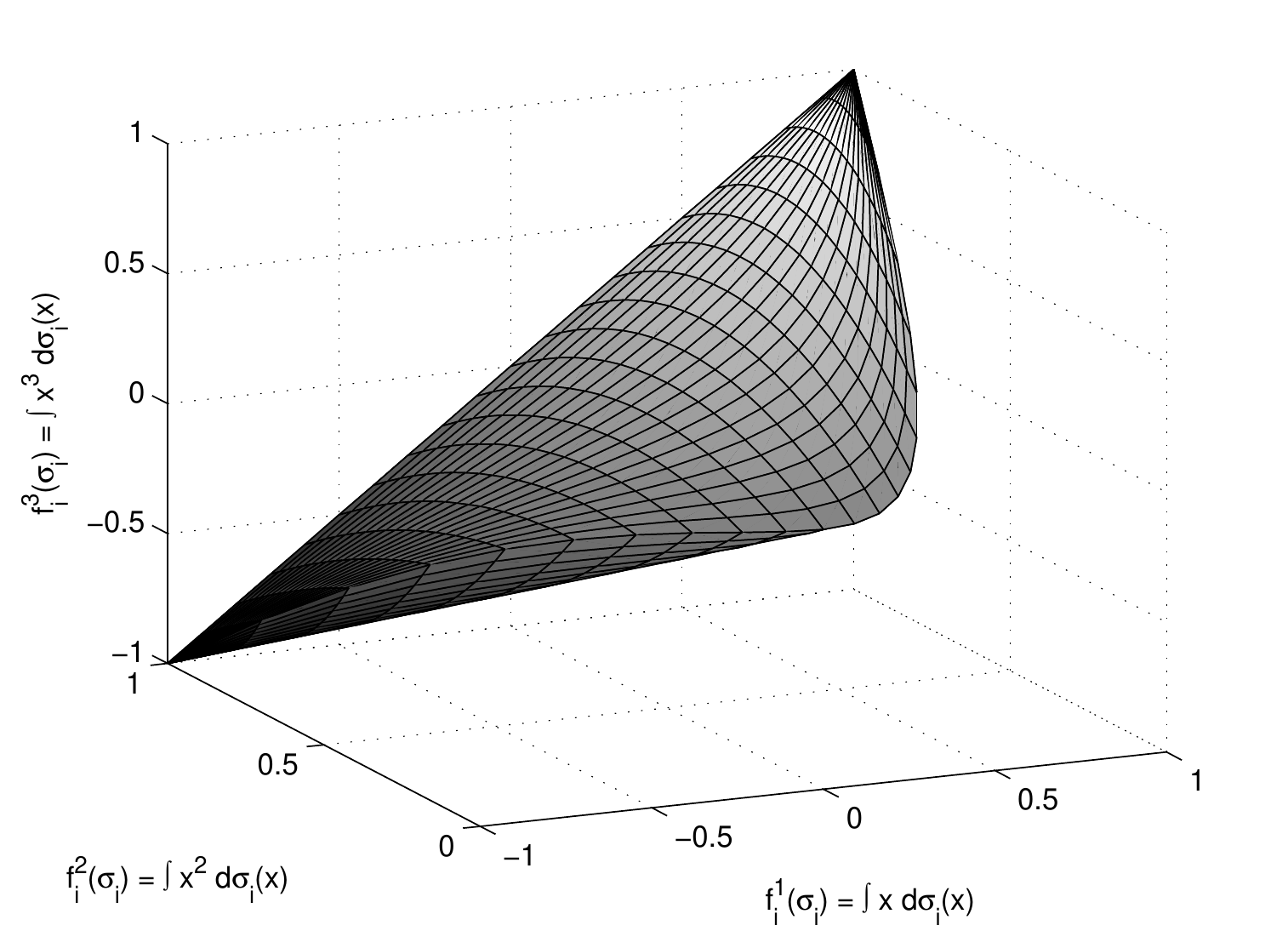}
    \caption{The space $f_i(\Delta_i) \cong \Delta_i/W_i$ of possible moments for either player's mixed strategy under the payoffs given in \eqref{eq:expayoffs} due to a measure $\sigma_i$ on $[-1,1]$.  The zeroth moment, which is identically unity, has been omitted.}
    \label{fig:momentspace}
\end{figure}

For each player the range of the indices in \eqref{eq:sepform} is
$0\leq j_i\leq 3$, so by Corollary \ref{cor:sepeq}, this game has an
equilibrium in which each player mixes among at most $4+1 = 5$ pure
strategies.  To produce this bound, we have not used any information
about the payoffs except for the degree of the polynomials. However, notice that
there is extra structure here to be exploited.  For example, $u_2$
depends on the expected value $\int x^2 d\sigma_1(x)$, but not on
$\int x d\sigma_1(x)$ or $\int x^3 d\sigma_1(x)$.  In particular, player $2$
is indifferent between the two strategies $\pm x$ of player $1$ for
all $x$, insofar as this choice does not affect his payoff (though
it does affect what strategy profiles are equilibria). In the
following section, we will present improved bounds on the number of
strategies played in an equilibrium which take these simplifications
into account in a systematic manner.
\end{example}
\restorecounter

\section{The Rank of a Continuous Game}
\label{sec:rank}

In this section, we introduce the notion of the {\it rank of a
continuous game}. We use this notion to provide improved bounds on
the cardinality of the support of the equilibrium strategies for a
separable game. We also provide a characterization theorem which
states that a continuous game has finite rank if and only if it is
separable, thus showing that separable games are the largest class
of continuous games to which low-rank arguments apply (see subsection
\ref{subsec:rankchar}). We conclude this section by showing how to
compute the rank of a separable game in Subsection
\ref{subsec:rankcomp}, which may be read independently of the
section on the characterization theorem.

Our main result on bounding the support of equilibrium strategies
generalizes the following theorem of Lipton et al.\
\cite{lmm:plgss} to arbitrary separable games, thereby removing the two-player assumption and weakening the restriction that the strategy spaces be finite. The proof given in \cite{lmm:plgss} is essentially an algorithmic version of
Carath\'{e}odory's theorem. Here we provide a shorter nonalgorithmic
proof which illustrates some of the ideas to be used in establishing
the extended version of the theorem.

\begin{theorem}[Lipton et al.\ \cite{lmm:plgss}]
\label{thm:lmmrank} Consider a two-player finite (i.e.\ bimatrix)
game defined by matrices $R$ and $C$ of payoffs to the row and
column player, respectively. Let $\sigma$ be a Nash equilibrium of
the game. Then there exists a Nash equilibrium $\tau$ which yields
the same payoffs to both players as $\sigma$, but in which the
column player mixes among at most $\rank R + 1$ pure strategies and
the row player mixes among at most $\rank C + 1$ pure strategies.
\end{theorem}

\begin{proof}
Let $r$ and $c$ be probability column vectors corresponding to the mixed
strategies of the row and column players in the given equilibrium
$\sigma$.  Then the payoffs to the row and column players are $r'Rc$
and $r'Cc$.  Since $c$ is a probability vector, we can view $Rc$ as
a convex combination of the columns of $R$. These columns all lie in
the column span of $R$, which is a vector space of dimension $\rank
R$.  By Carath\'{e}odory's theorem, we can therefore write any
convex combination of these vectors using only $\rank R + 1$ terms \cite{bno:convex}.
That is to say, there is a probability vector $d$ such that $Rd =
Rc$, $d$ has at most $\rank R + 1$ nonzero entries, and a component
of $d$ is nonzero only if the corresponding component of $c$ was
nonzero.

Since $r$ was a best response to $c$ and $Rc = Rd$, $r$ is a best
response to $d$.  On the other hand, since $(r,c)$ was a Nash
equilibrium $c$ must have been a mixture of best responses to $r$.
But $d$ only assigns positive probability to strategies to which $c$
assigned positive probability.  Thus $d$ is a best response to $r$,
so $(r,d)$ is a Nash equilibrium which yields the same payoffs to
both players as $(r,c)$, and $d$ only assigns positive probability
to $\rank R + 1$ pure strategies.  Applying the same procedure to
$r$ we could find an $s$ which only assigns positive probability to
$\rank C + 1$ pure strategies and such that $(s,d)$ is a Nash
equilibrium with the same payoffs to both players as $(r,c)$.
\end{proof}

We will show in the following subsection that our extension of this
theorem provides a slightly tighter bound of $\rank R$ instead
of $\rank R + 1$ (respectively for $C$) in some cases, depending on
the structure of $R$ and $C$. This can be seen directly from the
proof given here. We considered convex combinations of the columns
of $R$ and noted that these all lie in the column span of $R$.  In
fact, they all lie in the affine hull of the set of columns of $R$.
If this affine hull does not include the origin, then it will have
dimension $\rank R - 1$.  The rest of the proof goes through using
this affine hull instead of the column span, and so in this case we
get a bound of $\rank R$ on the number of strategies played with
positive probability by the row player. Alternatively, the dimension
of the affine hull can be computed directly as the rank of the
matrix given by subtracting a fixed column of $R$ from all the other
columns of $R$.

\subsection{Bound on the Support of Equilibrium Strategies}
\label{subsec:rankdef}

By comparing the two-player case of \eqref{eq:sepform} with the
singular value decomposition for matrices, separable games can be
thought of as games of ``finite rank.''  Now we will define the rank
of a continuous game precisely and use it to give a bound on the
cardinality of the support of equilibrium strategies, generalizing Corollary \ref{cor:sepeq} and
Theorem \ref{thm:lmmrank}.  The primary tool will be the notion of
almost payoff equivalence from Definition \ref{def:equivrels}.  In
what follows, the dimension of a set will refer to the dimension of
its affine hull.

\begin{definition} The \textbf{rank} of a continuous game is defined to
be the $n$-tuple $\rho = (\rho_1,\ldots, \rho_n)$ where $\rho_i = \dim
\Delta_i/Y_i$. A game is said to have \textbf{finite rank} if
$\rho_i<\infty$ for all $i$.
\end{definition}

Since $W_i\subseteq Y_i$ and $V_i/W_i$ is finite-dimensional for any
separable game (see the proof of Theorem \ref{thm:seppayoffequiv}),
it is clear that separable games have finite rank, and furthermore
that $\rho_i \leq m_i$ for all $i$.  In subsection
\ref{subsec:rankchar} we show that all games of finite rank are also
separable, so the two conditions are equivalent.  Unlike the $m_i$,
the rank is defined for all continuous games and is
representation-independent.

We define the rank of a game in terms of almost payoff equivalence of mixed strategies by means of the subspace $Y_i$.  This notion of equivalence between two strategies for a given player means that no matter what the other players do, their payoffs will not enable them to distinguish these two strategies. The use of this equivalence relation reflects the fact that at a (mixed strategy) Nash equilibrium, the payoff to a player is equalized among all pure strategies which he plays with positive probability.  Therefore the player is free to switch to any other mixed strategy supported on the same pure strategies, as long as the change does not affect the other players, and the resulting strategy profile will remain a Nash equilibrium.


Using the rank of a game, Corollary \ref{cor:sepeq} and Theorem
\ref{thm:lmmrank} can now be improved as follows:

\begin{theorem}
\label{thm:seprank} Given an equilibrium $\sigma$ of a separable
game with rank $\rho$, there exists an equilibrium $\tau$ such that
each player $i$ mixes among at most $\rho_i + 1$ pure strategies and
$u_i(\sigma) = u_i(\tau)$.

If $\dim \Delta_i/X_i = 1$ and the
metric space $C_i$ is connected, then this bound can be improved so
that $\tau_i$ is a pure strategy.
\end{theorem}

\begin{proof}
By Theorem \ref{thm:seppayoffequiv}, we can assume without loss of
generality that each player's mixed strategy $\sigma_i$ is finitely
supported.  Fix $i$, let $\psi_i: V_i\rightarrow V_i/Y_i$ denote the
canonical projection transformation and let $\sigma_i = \sum_j
\lambda^j s_i^j$ be a finite convex combination of pure strategies.
By linearity of $\psi_i$ we have
\begin{equation*}
    \psi_i(\sigma_i) = \sum_j \lambda^j \psi_i(s_i^j).
\end{equation*}
Carath\'{e}odory's theorem states that (renumbering the $s_i^j$ and
adding some zero terms if necessary) we can write
\begin{equation*}
    \psi_i(\sigma_i) = \sum_{j = 0}^{\rho_i} \mu^j \psi_i(s_i^j),
\end{equation*}
a convex combination potentially with fewer terms.  Let $\tau_i =
\sum_{j = 0}^{\rho_i}\mu^j s_i^j$.  Then $\psi_i(\sigma_i) =
\psi_i(\tau_i)$. Since $\sigma$ was a Nash equilibrium, and
$\sigma_i$ is almost payoff equivalent to $\tau_i$, $\sigma_j$ is a
best response to $(\tau_i,\sigma_{-i,j})$ for all $j\neq i$.  On the
other hand $\sigma_i$ was a mixture among best responses to the
mixed strategy profile $\sigma_{-i}$, so the same is true of
$\tau_i$, making it a best response to $\sigma_{-i}$.  Thus
$(\tau_i,\sigma_{-i})$ is a Nash equilibrium.

If $\dim \Delta_i/X_i = 1$ and $C_i$ is connected, then $C_i/X_i$ is connected, compact, and one-dimensional, i.e.\ it is an interval.\  Therefore it is convex, so $\Delta_i/X_i = \conv(C_i/X_i) = C_i/X_i$.  This implies that there exists a pure strategy $s_i$ which is payoff equivalent to $\sigma_i$, so we may take $\tau_i = s_i$ and $(\tau_i,\sigma_{-i})$ is a Nash equilibrium.

Beginning with this equilibrium and repeating the above steps for each player in turn completes the construction of $\tau$ and the final statement of the theorem is clear.
\end{proof}

While the preceding theorem was the original reason for our choice of the
definition of $\rho$, the definition turns out to have other
interesting properties which we study below.  The following
alternative characterization of the rank of a continuous game is
more concrete than the definition given above.  This theorem
simplifies the proofs of many rank-related results and will be
applied to the problem of computing the rank of separable games in
Subsection \ref{subsec:rankcomp}.

\begin{theorem}
\label{thm:rankchar} The rank $\rho_i$ for player $i$ in a
continuous game is given by the smallest $r_i$ such that there exist
continuous functions $g_i^k: C_i\rightarrow\Rm$ and $h_{i,j}^k:
C_{-i}\rightarrow\Rm$ which satisfy
\begin{equation}
\label{eq:rankchar}
u_j(s) = h_{i,j}^0(s_{-i}) + \sum_{k = 1}^{r_i} g_i^k(s_i ) h_{i,j}^k(s_{-i})
\end{equation}
for all $s\in C$ and $j\neq i$ ($\rho_i = \infty$ if and only if no
such representation exists).  Furthermore, the minimum value of $r_i
= \rho_i$ is achieved by functions $g_i^k(s_i)$ of the form
$u_j(s_i,s_{-i})$ for some $s_{-i}\in C_{-i}$ and $j\neq i$ and
functions $h_{i,j}^k(s_{-i})$ of the form $\int
u_j(\cdot,s_{-i})d\sigma_i$ for some $\sigma_i\in V_i$.
\end{theorem}

\begin{proof}
Throughout the proof we will automatically extend any functions $g_i^k:C_i\rightarrow\Rm$ to all of $V_i$ in the canonical way.  Suppose we are given a representation of the form \eqref{eq:rankchar}.  Let $g_i: C_i\rightarrow\Rm^{r_i}$ be defined by $g_i(s_i) = \left(g_i^1(s_i),\ldots,g_i^{r_i}(s_i)\right)$.  By definition, $\rho_i$ is the dimension of $\Delta_i/Y_i$.  Let $Z_i$ denote the subspace of $V_i$ parallel to $\Delta_i$, i.e.\ the space of all signed measures $\sigma_i$ such that $\int \sigma_i = 0$.  Then $\rho_i = \dim Z_i/(Z_i\cap Y_i)$.  By \eqref{eq:rankchar} any signed measure which is in $Z_i$ and in $\kernel g_i$ is almost payoff equivalent to the zero measure, so $Z_i\cap \kernel g_i \subseteq Z_i\cap Y_i$ and therefore
\begin{equation*}
\rho_i = \dim Z_i/(Z_i\cap Y_i)\leq \dim Z_i/(Z_i\cap \kernel g_i) = \dim g_i(Z_i) \leq r_i.
\end{equation*}

It remains to show that if $\rho_i < \infty$ then there exists a representation of the form \eqref{eq:rankchar} with $r_i = \rho_i$.  Recall that $Y_i$ is defined to be
\begin{equation*}
Y_i = \bigcap_{\stackrel{j\neq i}{s_{-i}\in C_{-i}}} \kernel u_j(\cdot,s_{-i})
\end{equation*}
where $u_j(\cdot,s_{-i})$ is interpreted as a linear functional on
$V_i$.  Since $\rho_i = \dim Z_i / (Z_i\cap Y_i)$, we can choose
$\rho_i$ linear functionals, call them $g_i^1,\ldots, g_i^{\rho_i}$,
from the collection of functionals whose intersection forms $Y_i$,
such that $Z_i\cap Y_i = Z_i\cap \kernel g_i$, where $g_i =
(g_i^1,\ldots,g_i^{\rho_i})$ as above.  We cannot choose a smaller
collection of linear functionals and achieve $Z_i\cap Y_i =
Z_i\cap\kernel g_i$, because $\rho_i = \dim Z_i / (Z_i\cap Y_i)$.
Note that $Z_i\cap \kernel g_i = \kernel
\left(1,g_i^1,\ldots,g_i^{\rho_i}\right)$ where $1$ is the linear
functional $1(\sigma_i) = \int d\sigma_i$.  Therefore no functional
can be removed from the list $(1,g_i) =
(1,g_i^1,\ldots,g_i^{\rho_i})$ without affecting the kernel of the
transformation $(1,g_i)$, so the functionals
$1,g_i^1,\ldots,g_i^{\rho_i}$ are linearly independent.

This means that any of the linear functionals $u_j(\cdot,s_{-i})$
(the intersection of whose kernels yields $Y_i$) can be written
uniquely as a linear combination of the functionals
$1,g_i^1,\ldots,g_i^{\rho_i}$.  That is to say, there are unique
functions $h_{i,j}^k$ such that \eqref{eq:rankchar} holds with the
functions $g_i^k$ constructed here and $r_i=\rho_i$.  The $g_i^k$
are continuous by construction, so to complete the proof we must
show that the functions $h_{i,j}^k$ are continuous as well.  Since
the functionals $1,g_i^1,\ldots,g_i^{\rho_i}$ are linearly
independent, we can choose a measure $\sigma_i^k\in V_i$ which makes
the $k^{\text{th}}$ of these functionals evaluate to unity and all
the others zero.  Substituting these values into \eqref{eq:rankchar}
shows that $h_{i,j}^k(s_{-i})=\int u_j(\cdot,s_{-i})d\sigma_i^k$.
Since $u_j$ is continuous, $h_{i,j}^k$ is therefore also continuous.
\end{proof}

Note that in the statement of Theorem \ref{thm:rankchar} we have
distinguished the component $h_{i,j}^0(s_{-i})$ in $u_j$.  We have
shown that this distinction follows from the definition of $\rho_i$,
but there is also an intuitive game theoretic reason why this
separation is natural.\  As mentioned above, $\rho_i$ is intended to
capture the number of essential degrees of freedom that player $i$
has in his choice of strategy when playing a Nash equilibrium.
Theorems \ref{thm:seprank} and \ref{thm:rankchar} together
show that player $i$ only needs to take the other players' utilities
into account to compute this number, and not his own.  But player
$i$ is only concerned with the other players' utilities insofar as
his own strategic choice affects them.  The function
$h_{i,j}^0(s_{-i})$ captures the part of player $j$'s utility which
does not depend on player $i$'s strategy, so whether this function
is zero or not it has no effect on the rank $\rho_i$.

Also note that while Theorem \ref{thm:rankchar} gives decompositions of the utilities in terms of the $\rho_i$, it is not in general possible to summarize all these decompositions by writing the utilities in the form \eqref{eq:sepform} with $m_i = \rho_i$ for all $i$.  For a trivial counterexample, consider any game in which $u_i(s) = h_i^0(s_{-i})$ is independent of $s_i$ for all $i$.  Then Theorem \ref{thm:rankchar} implies that $\rho_i = 0$ for all $i$, but the utilities are nonzero so we cannot take $m_i = 0$ in \eqref{eq:sepform} for any $i$.

We close this subsection with an application.  If a submatrix is
formed from a matrix by ``sampling,'' i.e.\ selecting a subset of the
rows and columns, the rank of the submatrix is bounded by the rank
of the original matrix.  Theorem \ref{thm:rankchar} shows that the
same is true of continuous games, because a factorization of the
form \eqref{eq:rankchar} for a game immediately provides a
factorization for any smaller game produced by restricting the
players' choices of strategies.

\begin{corollary}
\label{cor:ranksamp}Let $(\{C_i\}, \{u_i\})$ be a continuous game
with rank $\rho$ and $\tilde{C}_i$ be a nonempty compact subset of
$C_i$ for each $i$, with $\tilde{u}_i = u_i \big|_{\tilde{C}}$. Let
$\tilde \rho$ denote the rank of the game $(\{\tilde{C}_i\},
\{\tilde{u}_i\})$. Then we have $\tilde{\rho}_i\leq \rho_i$ for all
$i$.
\end{corollary}

\begin{definition}The game $(\{\tilde{C}_i\}, \{\tilde{u}_i\})$ in
Corollary \ref{cor:ranksamp} is called a \textbf{sampled game} or a
\textbf{sampled version} of $(\{C_i\}, \{u_i\})$.
\end{definition}

Note that if we take $\tilde{C}_i$ to be finite for each $i$, then
the sampled game is a finite game.  If the original game is
separable and hence has finite rank, then Corollary
\ref{cor:ranksamp} yields a uniform bound on the complexity of
finite games which can arise from this game by sampling.  This fact
is applied to the problem of computing approximate equilibria in
Section \ref{sec:comp} below. Finally, note that there are other
kinds of bounds on the cardinality of the support of equilibria
(e.g., for special classes of polynomial games as studied by Karlin
\cite{karlin:tig}) which do not share this sampling property.

\subsection{Characterizations of separable games}
\label{subsec:rankchar}

In this section we present a characterization theorem for separable
games. We also provide an example that shows that the assumptions of
this theorem cannot be weakened.

\begin{theorem}
\label{thm:sepcond} For a continuous game, the following are
equivalent:
\begin{enumerate}
\item \label{item:sep} The game is separable.
\item \label{item:fr} The game has finite rank.
\item \label{item:fs} For each player $i$, every countably supported mixed strategy $\sigma_i$ is almost payoff equivalent to a finitely supported mixed strategy $\tau_i$ with $\supp(\tau_i)\subset \supp(\sigma_i)$.
\end{enumerate}
\end{theorem}

To prove that finite rank implies separability we repeatedly apply
Theorem \ref{thm:rankchar}.  The proof that the technical condition
(\ref{item:fs}) implies (\ref{item:fr}) uses a linear algebraic
argument to show that $\linspan C_i/Y_i$ is finite dimensional
and then a topological argument along the lines of the proof of
Theorem \ref{thm:seppayoffequiv} to show that $V_i/Y_i$ is also
finite dimensional.

 After the proof of Theorem \ref{thm:sepcond} we will give an explicit example of a game in which all mixed strategies are payoff equivalent to pure strategies, but for which the containment $\supp(\tau_i)\subset \supp(\sigma_i)$ in condition (\ref{item:fs}) fails.  In light of Theorem \ref{thm:sepcond} this will show that the constructed game is nonseparable and that the containment $\supp(\tau_i)\subset \supp(\sigma_i)$ cannot be dropped from condition (\ref{item:fs}), even if the other assumptions are strengthened.

\begin{proof}
(\ref{item:sep} $\Rightarrow$ \ref{item:fs}) This was proven in Theorem \ref{thm:seppayoffequiv}.

(\ref{item:sep} $\Rightarrow$ \ref{item:fr}) This follows from the proof of Theorem \ref{thm:seppayoffequiv}.

(\ref{item:fr} $\Rightarrow$ \ref{item:sep}) We will prove this by
induction on the number of players $n$.  When $n=1$ the
statement is trivial and the case $n=2$ follows immediately from
Theorem \ref{thm:rankchar}.  Suppose we have an $n$-player
continuous game with $\rho_i<\infty$ for all $i$ and that we have proven that $\rho_i<\infty$ for all $i$ implies separability for $(n-1)$-player games.  By fixing any
signed measure $\sigma_n\in V_n$ we can form an $(n-1)$-player
continuous game from the given game by removing the $n^{\text{th}}$
player and integrating all payoffs of players $i<n$ with respect to
$\sigma_n$, yielding a new game with payoffs $\tilde{u}_i(s_{-n}) =
\int u_i(s_n,s_{-n})d\sigma_n(s_n)$.

From the definition of $Y_i$, it is clear that $Y_i \subseteq
\tilde{Y}_i$ for all $1\leq i<n$. Therefore $\tilde{\rho}_i = \dim
\Delta_i/\tilde{Y}_i \leq \dim \Delta_i/Y_i = \rho_i < \infty$ for
$1\leq i<n$
 so the $(n-1)$-player game has finite rank.  By the induction hypothesis, that means that the function
$\tilde{u}_1$ is a separable function of the strategies
$s_1,\ldots,s_{n-1}$.  Theorem \ref{thm:rankchar} states that there exist continuous functions $g_n^k$ and $h_{n,1}^k$ such that
\begin{equation}
\label{eq:finiterankimpliessep}
u_1(s) = h_{n,1}^0(s_{-n}) + \sum_{k = 1}^{\rho_n} g_n^k(s_n) h_{n,1}^k(s_{-n})
\end{equation}
where $h_{n,1}^k = \int u_1(s)d\sigma_n^k$ for some $\sigma_n^k\in
V_n$.  Therefore by choosing $\sigma_n$ appropriately we can make
$\tilde{u}_1 = h_{n,1}^k$ for any $k$, so $h_{n,1}^k(s_{-n})$ is a
separable function of $s_1,\ldots,s_{n-1}$ for all $k$.  By
\eqref{eq:finiterankimpliessep} $u_1$ is a separable function of
$s_1,\ldots,s_n$.  The same argument works for all the $u_i$ so the
given game is separable and the general case is true by induction.

(\ref{item:fs} $\Rightarrow$ \ref{item:fr}) Let $\psi_i:V_i\rightarrow V_i/Y_i$ be the canonical projection transformation.  First we will prove that $\linspan \psi_i(C_i)$ is finite dimensional.\  It suffices to prove that for every countable subset $\tilde{C}_i = \{s_i^1, s_i^2,\ldots\}\subseteq C_i$, the set $\psi_i(\tilde{C}_i)$ is linearly dependent.  Let $\{p^k\}$ be a sequence of positive reals summing to unity.  Define the mixed strategy
\begin{equation*}
    \sigma_i = \sum_{k=1}^{\infty} p^k s_i^k.
\end{equation*}
By assumption there exists an $M$ and $q^1,\ldots, q^M\geq 0$ summing to unity such that
\begin{equation*}
\psi_i(\sigma_i) = \psi_i\left(\sum_{k=1}^M q^k s_i^k\right) = \sum_{k=1}^M q^k \psi_i(s_i^k).
\end{equation*}
Let $\alpha = \displaystyle\sum_{k=M+1}^{\infty} p^k > 0$ and define the mixed strategy
\begin{equation*}
    \tau_i = \sum_{k=M+1}^{\infty} \frac{p^k}{\alpha} s_i^k.
\end{equation*}
Applying the assumption again shows that there exist $N$ and
$r^{M+1},\ldots,r^N$ such that 
\begin{equation*}
 \psi_i(\tau_i) = \psi_i\left(\sum_{k=M+1}^N r^k s_i^k\right) = \sum_{k=M+1}^N r^k \psi_i(s_i^k).
\end{equation*}
Therefore
\begin{equation*}
\begin{split}
\sum_{k=1}^M p^k \psi_i(s_i^k) & = \psi_i\left(\sum_{k=1}^M p^k s_i^k\right) = \psi_i(\sigma_i-\alpha\tau_i) = \psi_i(\sigma_i) - \alpha\psi_i(\tau_i) \\ &= \sum_{k=1}^M q^k \psi_i(s_i^k) - \sum_{k=M+1}^N \alpha r^k\psi_i(s_i^k),
\end{split}
\end{equation*}
and rearranging terms shows that $\sum_{k=1}^M (p^k - q^k)\psi_i(s_i^k) + \sum_{k=M+1}^N \alpha r^k\psi_i(s_i^k) = 0$.  Also $\sum_{k=1}^M (p^k - q^k) = -\alpha < 0$, so $p^k - q^k \neq 0$ for some $k$.  Therefore $\psi_i(\tilde{C}_i)$ is linearly dependent, so $\linspan\psi_i(C_i)$ is finite dimensional.

Since $Y_i$ is closed, $V_i/Y_i$ is a Hausdorff topological vector
space under the quotient topology and $\psi_i$ is continuous with
respect to this topology \cite{r:fa}.  Being finite dimensional, the
subspace $\linspan\psi_i(C_i)\subseteq V_i/Y_i$ is also closed
\cite{r:fa}.  Thus we have
\begin{equation*}
V_i/Y_i = \psi_i(V_i) = \psi_i\left(\overline{\linspan C_i}\right) \subseteq\overline{\psi_i(\linspan C_i)} = \overline{\linspan \psi_i(C_i)} = \linspan\psi_i(C_i) \subseteq V_i/Y_i
\end{equation*}
where the first step is by definition, the second follows from
Proposition \ref{prop:standardtop}, the next two are by continuity
and linearity of $\psi_i$, and the final two are because
$\linspan\psi_i(C_i)$ is a closed subspace of $V_i/Y_i$.  Therefore $\rho_i =
\dim\Delta_i/Y_i\leq \dim V_i/Y_i = \dim
\linspan\psi_i(C_i)<\infty$.
\end{proof}
The following counterexample shows that the containment
$\supp{\tau_i}\subset\supp{\sigma_i}$ is a necessary part of
condition \ref{item:fs} in Theorem \ref{thm:sepcond} by showing that
there exists a nonseparable continuous game in which every mixed
strategy is payoff equivalent to a pure strategy.

\begin{example}
Consider a two-player game with $C_1 = C_2 = [0,1]^\omega$, the set of all infinite sequences of reals in $[0,1]$, which forms a compact metric space under the metric
\begin{equation*}
    d(x,x') = \sup_i \frac{\lvert x_i - x'_i \rvert}{i}.
\end{equation*}
Define the utilities
\begin{equation*}
    u_1(x,y) = u_2(x,y) = \sum_{i=1}^\infty 2^{-i}x_i y_i.
\end{equation*}
To show that this is a continuous game we must show that $u_1$ is continuous.  Assume $d(x,x'),d(y,y')\leq\delta$.  Then $\lvert x_i - x'_i \rvert \leq \delta i$ and $\lvert y_i - y'_i \rvert \leq \delta i$, so
\begin{equation*}
\begin{split}
    \lvert u_1(x,& y) - u_1(x',y') \rvert = \left\lvert \sum_{i=1}^\infty 2^{-i}(x_i y_i - x'_i y'_i)\right\rvert \\
    & = \left\lvert \sum_{i=1}^\infty 2^{-i} (x_i y_i - x'_i y_i + x'_i y_i - x'_i y'_i)\right\rvert \\
    & \leq \sum_{i=1}^\infty 2^{-i} \left( y_i\lvert x_i - x'_i\rvert + x'_i\lvert y_i - y'_i\rvert\right) \\
    & \leq \sum_{i=1}^\infty 2^{-i} (2\delta i) = \left(2\sum_{i=1}^\infty 2^{-i}i\right)\delta.
\end{split}
\end{equation*}
Thus $u_1 = u_2$ is continuous (in fact Lipschitz), making this a continuous game.

Let $\sigma$ and $\tau$ be mixed strategies for the two players.  By the Tonelli-Fubini theorem,
\begin{equation*}
u_1(\sigma,\tau) = \int u_1 d(\sigma\times\tau) = \sum_{i = 1}^\infty 2^{-i}\left(\int x_i d\sigma\right)\left(\int y_i d\tau\right).
\end{equation*}
Thus $\sigma$ is payoff equivalent to the pure strategy $\left(\int x_1 d\sigma,\int x_2 d\sigma,\ldots\right)\in C_1$ and similarly for $\tau$, so this game has the property that every mixed strategy is payoff equivalent to a pure strategy.

Finally we will show that this game is nonseparable.  Let $e^i\in C_1$ be the element having component $i$ equal to unity and all other components zero.  Let $\{p_i\}$ be a sequence of positive reals summing to unity and define the probability distribution $\sigma = \sum_{i=1}^\infty p_i e^i\in \Delta_1$.  Suppose $\sigma$ were almost payoff equivalent to some mixture among finitely many of the $e^i$, call it $\tau = \sum_{i = 1}^\infty q_i e^i$ where $q_i = 0$ for $i$ greater than some fixed $N$.  Let $e_{N+1}$ be the strategy for player $2$.  Then the payoff if player $1$ plays $\sigma$ is
\begin{equation*}
u_2(\sigma,e_{N+1}) = \int 2^{-(N+1)}x_{N+1}d\sigma = 2^{-(N+1)}p_{N+1}.
\end{equation*}
Similarly, if he chooses $\tau$ the payoff is $2^{-(N+1)}q_{N+1}$.  Since $p_{N+1}>0$ and $q_{N+1}=0$, this contradicts the assumption that $\sigma$ and $\tau$ are almost payoff equivalent.  Thus condition \ref{item:fs} of Theorem \ref{thm:sepcond} does not hold, so this game is not separable.

Therefore the condition that all mixed strategies be payoff
equivalent to finitely supported strategies does not imply
separability, even if a uniform bound on the size of the support is
assumed.  Hence the containment
$\supp{\tau_i}\subset\supp{\sigma_i}$ cannot be removed from
condition \ref{item:fs} of Theorem \ref{thm:sepcond}.
\end{example}

\subsection{Computing the rank of a separable game}
\label{subsec:rankcomp}

In this subsection we construct a formula for the rank of an arbitrary
separable game and then specialize it to get formulas for the ranks
of polynomial and finite games.  For clarity of presentation we
first prove a bound on the rank of a separable game which uses an
argument that is similar to but simpler than the argument for the
exact formula.  While it is possible to prove all the results in
this section directly from the definition $\rho_i = \dim
\Delta_i/Y_i$, we will give proofs based on the alternative
characterization in Theorem \ref{thm:rankchar} because they are
easier to understand and provide more insight into the structure of
the problem.

Given a separable game in the standard form \eqref{eq:sepform}, construct a matrix $S_{i,j}$ for players $i$ and $j$ which has $m_i$ columns and $\Pi_{k\neq i} m_k$ rows and whose elements are defined as follows.  Label each row with an $(n-1)$-tuple $(l_1,\ldots,l_{i-1},l_{i+1},\ldots,l_n)$ such that $1\leq l_k\leq m_k$; the order of the rows is irrelevant.  Label the columns $l_i = 1,\ldots, m_i$.  Each entry of the matrix then corresponds to an $n$-tuple $(l_1,\ldots, l_n)$.  The entry itself is given by the coefficient $a_j^{l_1 \cdots l_n}$ in the utility function $u_j$.

Let $f_i(s_i)$ denote the column vector whose components are $f_i^1(s_i),\ldots,f_i^{m_i}(s_i)$ and $f_{-i}(s_{-i})$ denote the row vector whose components are the products $f_1^{l_1}(s_1)\cdots f_{i-1}^{l_{i-1}}(s_{i-1})f_{i+1}^{l_{i+1}}(s_{i+1})\cdots f_n^{l_n}(s_n)$ ordered in the same way as the $(n-1)$-tuples $(l_1,\ldots,l_{i-1},l_{i+1},\ldots,l_n)$ were ordered above.  Then $u_j(s) = f_{-i}(s_{-i})S_{i,j} f_i(s_i)$.

\savecounter{exthreeplayer}
\begin{example}
\label{example:threeplayerpoly}
We introduce a new example game to clarify the subtleties of computing ranks when there are more than two players; we will return to Example \ref{example:ex1} later.  Consider the three player polynomial game with strategy spaces $C_1 = C_2 = C_3 = [-1,1]$ and payoffs
\begin{equation}
\label{eq:exthreeplayer}
\begin{split}
u_1(x,y,z) = 1 & + 2x + 3x^2 + 2yz + 4xyz + 6x^2 yz \\
& + 3y^2 z^2 + 6xy^2 z^2 + 9x^2 y^2 z^2 \\
u_2(x,y,z) = 7 & + 2x + 3x^2 + 2y + 4xy + 6x^2 y \\
& + 3z^2 + 6xz^2 + 9x^2 z^2 \\
u_3(x,y,z) = -z & - 2xz - 3x^2z - 2yz - 4xyz - 6x^2 yz \\
& - 3yz^2 - 6xyz^2 - 9x^2 yz^2 \\
\end{split}
\end{equation}
where $x$, $y$, and $z$ are the strategies of player $1$, $2$, and
$3$, respectively.  Order the functions $f_k^l$ so that  $f_1(x) =
\begin{bmatrix}1& x &x^2\end{bmatrix}'$ and similarly for $f_2$ and
$f_3$ with $x$ replaced by $y$ and $z$, respectively.  If we wish to
write down the matrices $S_{1,2}$ and $S_{1,3}$ we must choose an
order for the pairwise products of the functions $f_2^l$ and
$f_3^l$.  Here we will choose the order $f_{-1}(y,z) =
\begin{bmatrix} 1& y & y^2 & z & yz & y^2z & z^2 & yz^2&
y^2z^2\end{bmatrix}$.  We can write down the desired matrices
immediately from the given utilities.
\begin{equation*}
S_{1,2} = \begin{bmatrix}7 & 2  & 3 \\ 2 & 4 & 6 \\ 0 & 0 & 0 \\ 0 & 0 & 0 \\ 0 & 0 & 0\\ 0 & 0 & 0 \\ 3 & 6 & 9 \\ 0 & 0 & 0 \\ 0 & 0 & 0\end{bmatrix}
\text{, }
S_{1,3} = \begin{bmatrix}0 & 0 & 0 \\ 0 & 0 & 0 \\ 0 & 0 & 0 \\ -1 & -2  & -3 \\ -2 & -4 & -6 \\ 0 & 0 & 0 \\ 0 & 0 & 0 \\  -3 & -6 & -9 \\ 0 & 0 & 0\end{bmatrix}
\end{equation*}
This yields $u_2(x,y,z) = f_{-1}(y,z)S_{1,2}f_1(x)$ and $u_3(x,y,z) = f_{-1}(y,z)S_{1,3}f_1(x)$ as claimed.
\end{example}

Define $S_i$ to be the matrix with $m_i$ columns and $(n-1)\Pi_{j\neq i}m_j$ rows which consists of all the matrices $S_{i,j}$ for $j\neq i$ stacked vertically (in any order).  In the example above, $S_1$ would be the $18 \times 3$ matrix obtained by placing $S_{1,2}$ above $S_{1,3}$ on the page.

\begin{theorem}
\label{thm:rankbound} The rank of a separable game is bounded by
$\rho_i\leq \rank S_i$.
\end{theorem}

\begin{proof}
Using any of a variety of matrix factorization techniques (e.g.\ the
singular value decomposition), we can write $S_i$ as
\begin{equation*}
S_i = \sum_{k=1}^{\rank S_i} v^k w^k
\end{equation*}
for some column vectors $v^k$ and row vectors $w^k$. The vectors
$v^k$ will have length $(n-1)\Pi_{j\neq i}m_j$ since that is the
number of rows of $S_i$.  Because of the definition of $S_i$, we can
break each $v^k$ into $n-1$ vectors of length $\Pi_{j\neq i}m_j$,
one for each player except $i$, and let $v_j^k$ be the vector
corresponding to player $j$.  Then we have
\begin{equation*}
S_{i,j} = \sum_{k=1}^{\rank S_i} v_j^k w^k
\end{equation*}
for all $j\neq i$.  Define the linear combinations $g_i^k(s_i) = w^k
f_i(s_i)$ and $h_{i,j}^k = f_{-i}(s_{-i})v_j^k$, which are obviously
continuous functions.  Then
\begin{equation*}
u_j(s) = f_{-i}(s_{-i})S_{i,j} f_i(s_i) = \sum_{k=1}^{\rank S_i} g_i^k(s_i)h_{i,j}^k(s_{-i})
\end{equation*}
for all $s\in C$ and $j\neq i$, so $\rho_i \leq \rank S_i$ by Theorem \ref{thm:rankchar}.
\end{proof}

\begin{example}
To demonstrate the power of the bound in Theorem \ref{thm:rankbound}
we will use it to give an immediate proof of Theorem
\ref{thm:lmmrank}.  Consider any two-player finite game, where the
first player chooses rows and the second player chooses columns. Let
$C_i = \{1,\ldots, m_i\}$ for $i = 1,2$ and let $R$ and $C$ be the
matrices of payoffs to the row and column players, respectively. We
can then define $f_i^l(s_i)$ to be unity if $s_i = l$ and zero
otherwise.  This gives
\begin{equation*}
\begin{split}
u_1(s_1,s_2) & = f_1(s_1)' R f_2(s_2) \\
u_2(s_1,s_2) & = f_2(s_2) C' f_1(s_1)
\end{split}
\end{equation*}
so $S_1 = C'$ and $S_2 = R$.  Therefore by Theorem
\ref{thm:rankbound}, $\rho_1\leq \rank S_1 = \rank C$ and $\rho_2
\leq \rank S_2 = \rank R$.  Substituting these bounds into Theorem
\ref{thm:seprank} yields Theorem \ref{thm:lmmrank}, so we have in
fact generalized the results of Lipton et al.\ \cite{lmm:plgss}.
\end{example}

It is easy to see that there are cases in which the bound in Theorem
\ref{thm:rankbound} is not tight.  For example, this will be the
case (for generic coefficients $a_i^{j_1\cdots j_n}$) if $m_i \geq
2$ for each $i$ and $f_i^k$ is the same function for all $k$.

Fortunately we can use a technique similar to the one used above to
compute $\rho_i$ exactly instead of just computing a bound.  To do
so we need to write the utilities in a special form.  First we add the
new function $f_i^1(s_i) \equiv 1$ to the list of functions for
player $i$ appearing in the separable representation of the game if
this function does not already appear, relabeling the other $f_i^k$
as necessary.  Next we consider the set of functions $\{f_j^k\}$ for
each player $j$ in turn and choose a maximal linearly independent
subset.  For players $j\neq i$ any such subset will do; for player
$i$ we must include the function which is identically unity in the
chosen subset.  Finally we rewrite the utilities in terms of these
linearly independent sets of functions.  This is possible because
all of the $f_j^k$ are linear combinations of those which appear in
the maximal linearly independent sets.

From now on we will assume the utilities are in this form and that
$f_i^1(s_i)\equiv 1$.  Let $\overline{S}_{i,j}$ and $\overline{S}_i$
be the matrices $S_{i,j}$ and $S_i$ defined above, where the bar
denotes the fact that we have put the utilities in this special
form.  Let $T_i$ be the matrix $\overline{S}_i$ with its first
column removed.  Note that this column corresponds to the function
$f_i^1(s_i)\equiv 1$ which we have distinguished above, and
therefore represents the components of the utilities of players
$j\neq i$ which do not depend on player $i$'s choice of strategy. As
mentioned in the note following Theorem \ref{thm:rankchar}, these
components don't affect the rank.  This is exactly the reason that
we must remove the first column from $\overline{S}_i$ in order to
compute $\rho_i$.  We will prove that $\rho_i = \rank T_i$, but
first we need a lemma.

\begin{lemma}
\label{lem:linindprod} If the functions $f_j^1(s_j), \ldots, f_j^{m_j}(s_j)$
are linearly independent for all $j$, then the set of all
$\Pi_{j=1}^n m_j$ product functions of the form $f_1^{k_1}(s_1)\cdots f_n^{k_n}(s_n)$
is a linearly independent set.
\end{lemma}

\begin{proof}
It suffices to prove this in the case $n=2$, because the general
case follows by induction.  We prove the $n=2$ case by
contradiction. Suppose the set were linearly dependent.  Then there
would exist $\lambda_{k_1 k_2}$ not all zero such that
\begin{equation}
\label{eq:linindlemma}
\sum_{k_1=1}^{m_1}\sum_{k_2=1}^{m_2} \lambda_{k_1 k_2} f_1^{k_1}(s_1) f_2^{k_2}(s_2) = 0
\end{equation}
for all $s\in C$.  Choose $l_1$ and $l_2$ such that $\lambda_{l_1
l_2}\neq 0$.  By the linear independence assumption there exists a finitely supported
signed measure $\sigma_2$ such that $\int f_2^k d\sigma_2$ is unity
for $k = l_2$ and zero otherwise.  Integrating
\eqref{eq:linindlemma} with respect to $\sigma_2$ yields
\begin{equation*}
\sum_{k_1 = 1}^{m_1}\lambda_{k_1 l_2} f_1^{k_1}(s_1) = 0,
\end{equation*}
contradicting the linear independence assumption for
$f_1^1,\ldots,f_1^{m_1}$.
\end{proof}

\begin{theorem}
\label{thm:rankcomp} If the representation of a separable game
satisfies $f_i^1 \equiv 1$ and the set $\{f_j^1,\ldots,f_j^{m_j}\}$
is linearly independent for all $j$ then the rank of the game is
$\rho_i = \rank T_i$.
\end{theorem}

\begin{proof}
The proof that $\rho_i\leq\rank T_i$ follows essentially the same
argument as the proof of Theorem \ref{thm:rankbound}.  We use the
singular value decomposition to write $T_i$ as
\begin{equation*}
T_i = \sum_{k=1}^{\rank T_i} v^k w^k
\end{equation*}
for some column vectors $v^k$ and row vectors $w^k$. The vectors
$v^k$ will have length $(n-1)\Pi_{j\neq i}m_j$ since that is the
number of rows of $S_i$.  Let $v^0$ be the first column of
$\overline{S}_i$, which was removed from $\overline{S}_i$ to form
$T_i$.  Because of the definition of $T_i$ and $\overline{S}_i$, we
can break each $v^k$ into $n-1$ vectors of length $\Pi_{j\neq
i}m_j$, one for each player except $i$, and let $v_j^k$ be the
vector corresponding to player $j$.  Putting these definitions
together we get
\begin{equation*}
\overline{S}_{i,j} = v_j^0\begin{bmatrix}1 & 0 & \cdots & 0\end{bmatrix} + \sum_{k=1}^{\rank T_i} v_j^k \begin{bmatrix} 0 & w^k\end{bmatrix}.
\end{equation*}

Define the linear combinations $g_i^k(s_i) = \begin{bmatrix}0 &
w^k\end{bmatrix} f_i(s_i)$ and $h_{i,j}^k(s_{-i}) = f_{-i}(s_{-i})v_j^k$,
which are obviously continuous functions.  Then
\begin{equation*}
u_j(s) = f_{-i}(s_{-i})\overline{S}_{i,j} f_i(s_i) = h_{i,j}^0(s_{-i}) + \sum_{k=1}^{\rank T_i} g_i^k(s_i)h_{i,j}^k(s_{-i})
\end{equation*}
for all $s\in C$ and $j\neq i$, so $\rho_i \leq \rank T_i$ by Theorem \ref{thm:rankchar}.

To prove the reverse inequality, choose continuous functions $g_i^k(s_i)$ and $h_{i,j}^k(s_{-i})$ such that
\begin{equation*}
u_j(s) = h_{i,j}^0(s_{-i}) + \sum_{k=1}^{\rho_i} g_i^k(s_i)h_{i,j}^k(s_{-i})
\end{equation*}
holds for all $s\in C$ and $j\neq i$.  By Theorem \ref{thm:rankchar}
we can choose these so that $g_i^k(s_i)$ is of the form
$u_j(s_i,s_{-i})$ for some $s_{-i}\in C_{-i}, j\neq i$ and
$h_{i,j}^k(s_{-i})$ is of the form $\int u_j(\cdot,s_{-i})d\sigma_i$
for some $\sigma_i\in V_i$.  Substituting these conditions into
equation \eqref{eq:sepform} defining the form of a separable game
shows that $g_i^k(s_i) = w^k f_i(s_i)$ for some row vectors $w^k$
and $h_{i,j}^k = f_{-i}(s_{-i})v_j^k$ for some column vectors
$v_j^k$.  Define $w^0 = \begin{bmatrix}1 & 0 & \cdots &
0\end{bmatrix}$.  Then
\begin{equation*}
u_j(s) = \sum_{k=0}^{\rho_i} f_{-i}(s_{-i})'v_j^k w^k f_i(s_i)
\end{equation*}
for all $s\in C$ and $j\neq i$.

This expresses $u_j(s)$ as a linear combination of products of the
form $f_1^{k_1}(s_1)\cdots f_n^{k_n}(s_n)$.  By assumption the
sets $\{f_j^1,\ldots,f_j^{m_j}\}$ are linearly independent for all
$j$, and therefore the set of products of the form
$f_1^{k_1}(s_1)\cdots f_n^{k_n}(s_n)$ is linearly independent by
Lemma \ref{lem:linindprod}.  Thus the expression of $u_j(s)$ as a
linear combination of these products is unique.

But we also have $u_j(s) =
f_{-i}(s_{-i})'\overline{S}_{i,j}f_i(s_i)$ by definition of
$\overline{S}_{i,j}$, so uniqueness implies that $\overline{S}_{i,j}
= \sum_{k=0}^{\rho_i} v_j^k w^k$.  Let $v^k$ be the vector of length
$(n-1)\Pi_{j\neq i}m_j$ formed by concatenating the $v_j^k$ in the
obvious way.  Then $\overline{S}_i = \sum_{k=0}^{\rho_i} v^k w^k$.
Let $\tilde{w}^k$ be $w^k$ with its first entry removed.  By
definition of $T_i$ we have $T_i = \sum_{k=0}^{\rho_i} v^k
\tilde{w}^k$.  But $w^0$ is the standard unit vector with a $1$ in
the first coordinate, so $\tilde{w}^0$ is the zero vector and we may
therefore remove the $k=0$ term from the sum.  Thus $T_i =
\sum_{k=1}^{\rho_i} v^k \tilde{w}^k$, which proves that $\rank T_i
\leq \rho_i$.
\end{proof}

As corollaries of Theorem \ref{thm:rankcomp} we obtain formulas for the ranks of polynomial and finite games.
\begin{corollary}
Consider a game with polynomial payoffs
\begin{equation}
\label{eq:polyform}
u_i(s) = \sum_{j_1=0}^{m_1-1} \cdots \sum_{j_n=0}^{m_n-1} a_i^{j_1\cdots j_n} s_1^{j_1}\cdots s_n^{j_n}
\end{equation}
and compact strategy sets $C_i\subset\Rm$ which satisfy the cardinality condition $\left|C_i\right| > m_i$ for all $i$.  Then $T_i$ is $S_i$ with its first column removed and $\rho_i = \rank T_i$.
\end{corollary}

\begin{proof}
Linear independence of the $f_i^l$ follows from the cardinality condition and we have $f_i^0 \equiv 1$, so Theorem \ref{thm:rankcomp} applies.
\end{proof}

\usesavedcounter{ex1}
\begin{example}[cont'd]
Applying this formula to the utilities in \eqref{eq:expayoffs} shows that $\rho_1 = 1$ and $\rho_2 = 3$.
\end{example}
\restorecounter

\usesavedcounter{exthreeplayer}
\begin{example}[cont'd]
Applying this formula to the utilities in \eqref{eq:exthreeplayer}
shows that in this case $\rho_1 = 1$ and $\rho_2 = \rho_3 = 2$.
\end{example}
\restorecounter

\begin{corollary}
Consider an $n$-player finite game with strategy sets $C_i =
\{1,\ldots,m_i\}$ and payoff $a_i^{s_1 \cdots s_n}$ to player $i$ if
the players play strategy profile $(s_1,\ldots,s_n)$.  The utilities
can be written as
\begin{equation*}
u_i(s) = \sum_{j_1=1}^{m_1} \cdots \sum_{j_n=1}^{m_n} a_i^{j_1\cdots j_n} f_1^{j_1}(s_1)\cdots f_n^{j_n}(s_n)
\end{equation*}
where $f_i^l(s_i)$ is unity if $s_i = l$ and zero otherwise.  Let
$S_i$ be the matrix for player $i$ as defined above and let
$c_1,\ldots,c_{m_i}$ be the columns of $S_i$.  Then we may take $T_i
= \begin{bmatrix} c_2 - c_1 & \cdots & c_{m_i} - c_1\end{bmatrix}$
and $\rho_i = \rank T_i$.
\end{corollary}

\begin{proof}
If we replace $f_i^1$ with the function which is identically unity
then the linear independence assumption on the $f_k^l$ will still be
satisfied, so we can apply Theorem \ref{thm:rankcomp}.  After this
replacement, the coefficients in the new separable representation
for the game are
\begin{equation*}
\overline{a}_k^{j_1 \cdots j_n} =
\begin{cases}
a_k^{j_1 \cdots j_n} & \text{if $j_i = 1$,} \\
a_k^{j_1 \cdots j_n} - a_k^{j_1 \cdots j_{i-1} 1 j_{i+1} \cdots j_n} & \text{if $j_i \neq 1$.}
\end{cases}
\end{equation*}
Therefore if $c_1,\ldots, c_{m_i}$ are the columns of $S_i$ from the
original representation of the game we get $\overline{S}_i =
\begin{bmatrix}c_1 & c_2 - c_1 & \cdots & c_{m_i} -
c_1\end{bmatrix}$, so $T_i$ is as claimed and an application of
Theorem \ref{thm:rankcomp} completes the proof.
\end{proof}


\section{Computation of Nash Equilibria and Approximate Equilibria}
\label{sec:comp}

In this section, we study computation of exact and approximate Nash
equilibria. We first present an optimization formulation for the
computation of (exact) Nash equilibria of general separable games.
We show that for two-player polynomial games, this formulation has a
biaffine objective function and linear matrix inequality
constraints. We then present an algorithm for computing approximate
equilibria of two-player separable games with infinite strategy sets
which follows directly from the results on the rank of games given
in Section \ref{sec:rank} and compare it with known algorithms for
finite games.

\subsection{Computing Nash equilibria}
\label{subsec:compnash}

The moments of an equilibrium can in principle be computed by
nonlinear programming techniques using the following generalization
of the Nash equilibrium formulation presented by Ba\c{s}ar and Olsder
\cite{bo:dngt}:
\begin{proposition}
The following optimization problem has optimal value zero and the variables $x$ in any optimal solution are the moments of a Nash equilibrium strategy profile with payoff $p_i$ to player $i$:
\begin{equation*}
\begin{array}{cc}
\text{max}  & \sum_{i=1}^n \left[v_i(x) - p_i\right] \\
\text{s.t.} & x_i \in \Delta_i/W_i = f_i(\Delta_i)\text{ for all }i \\
            & v_i(f_i(s_i),x_{-i})\leq p_i\text{ for all }i\text{, all }s_i\in C_i \\
\end{array}
\end{equation*}
The function $f_i$ is the moment function defined in \eqref{eq:momentfunction} and $v_i$ is the payoff function on the moment spaces defined by $v_i\left(f_1(\sigma_1),\ldots,f_n(\sigma_n)\right) = u_i(\sigma)$.
\end{proposition}

\begin{proof} The constraints imply that $v_i(x) - p_i \leq 0$ for all $i$, so the objective function is
bounded above by zero.  Given any $n$-tuple of moments $x$ which form a Nash equilibrium, let $p_i = v_i(x)$ for all $i$.  Then the objective function evaluates to zero and all the constraints are satisfied, by definition of a Nash equilibrium.  Therefore the optimal objective function value is zero and it is attained at all Nash equilibria.

Conversely suppose some feasible $x$ and $p$ achieve objective
function value zero. Then the condition $v_i(x) - p_i \leq 0$
implies that $v_i(x) = p_i$ for all $i$.  Also, the final constraint
implies that player $i$ cannot achieve a payoff of more than $p_i$
by unilaterally changing his strategy.  Therefore the moments $x$
form a Nash equilibrium.
\end{proof}

To compute equilibria by this method, we require an explicit
description of the spaces of moments $\Delta_i/W_i$.  We also
require a method for describing the payoff $p_i$ to player $i$ if he plays
a best response to an $m_{-i}$-tuple of moments for the other
players.

While it seems doubtful that such descriptions could be found for
arbitrary $f_i^j$, they do exist for two-player polynomial games in
which the pure strategy sets are intervals.  In this case they can
be written in terms of linear matrix inequalities as in Parrilo's
treatment of the zero-sum case \cite{pp:polygames}.  This yields a
problem with biaffine objective and linear matrix inequality
constraints.

\usesavedcounter{ex1}
\begin{example}[cont'd]
Directly solving this nonconvex problem with MATLAB's \texttt{fmincon} has proven error-prone, as there appear to be many local minima which are not global.\  However, we were able to compute the equilibrium measures
\begin{equation*}
\begin{split}
  \sigma_1 &= 0.5532 \delta(x + 1) + 0.4468 \delta(x-0.1149), \\
  \sigma_2 &= \delta(y-0.7166)
\end{split}
\end{equation*}
(i.e.\ player $1$ plays the pure strategy $x = -1$ with probability $0.5532$ and so on) for the payoffs in \eqref{eq:expayoffs} by this method.
\end{example}
\restorecounter

\subsection{Computing $\epsilon$-equilibria}
\label{subsec:compeps} The difficulties in computing equilibria by
general nonconvex optimization techniques suggest the need for more
specialized systematic methods.  As a step toward this, we present
an algorithm for computing approximate Nash equilibria of two-player separable
games.  There are several possible definitions of approximate
equilibrium, but here we will use:

\begin{definition}
\label{def:epseq}A mixed strategy profile $\sigma\in\Delta$ is an \textbf{$\epsilon$-equilibrium} ($\epsilon \geq 0$) if
\begin{equation*}
    u_i(s_i,\sigma_{-i})\leq u_i(\sigma) + \epsilon
\end{equation*}
for all $s_i\in C_i$ and $i = 1,\ldots, n$.
\end{definition}

For $\epsilon=0$, the definition of an $\epsilon$-equilibrium reduces to that of a Nash
equilibrium. We consider computing an $\epsilon$-equilibrium of
separable games that satisfy the following assumption:

\begin{assumption}
\label{assumption:essential}
\
\begin{itemize}
\item There are two players.
\item The game is separable.
\item The utilities can be evaluated efficiently.
\end{itemize}
\end{assumption}

To simplify the presentation of our algorithm, we also adopt the
following assumption:

\begin{assumption}
\label{assumption:simplifying}
\
\begin{itemize}
\item The strategy spaces are $C_1 = C_2 = [-1,1]$.
\item The utility functions are Lipschitz.
\end{itemize}
\end{assumption}

In the description of the algorithm we will emphasize why Assumption
\ref{assumption:essential} is needed for our analysis.  After
presenting the algorithm we will discuss how Assumption
\ref{assumption:simplifying} could be relaxed.

\begin{theorem}
\label{thm:epseq}For $\epsilon>0$, the following algorithm computes an $\epsilon$-equilibrium of a game of rank $\rho$ satisfying Assumptions \ref{assumption:essential} and \ref{assumption:simplifying} in time polynomial in $\frac{1}{\epsilon}$ for fixed $\rho$ and time polynomial in the components of $\rho$ for fixed $\epsilon$ (for the purposes of asymptotic analysis of the algorithm with respect to $\rho$ the Lipschitz condition is assumed to be satisfied uniformly by the entire class of games under consideration).
\end{theorem}

\begin{algorithm}
\label{alg:epseq}
By the Lipschitz assumption there are real numbers $L_1$ and $L_2$ such that
\begin{equation*}
    \lvert u_i(s_i,s_{-i}) - u_i(s'_i,s_{-i})\rvert \leq L_i\lvert s_i - s'_i \rvert
\end{equation*}
for all $s_{-i}\in C_{-i}$ and $i = 1, 2$.  Clearly this is equivalent to requiring the same inequality for all $\sigma_{-i}\in\Delta_{-i}$.  Divide the interval $C_i$ into equal subintervals of length no more than $2\frac{\epsilon}{L_i}$; at most $\lceil \frac{L_i}{\epsilon}\rceil$ such intervals are required.  Let $\tilde{C}_i$ be the set of center points of these intervals, and construct a finite sampled game by restricting the strategy sets to the $\tilde{C}_i$.  Call the resulting payoff matrices $U_1$ and $U_2$.  Compute a Nash equilibrium of the sampled game.  To do so, iterate over all pairs of nonempty subsets $S_1\subseteq\tilde{C}_1$ and $S_2\subseteq\tilde{C}_2$ such that the cardinality of $S_i$ is at most $\rho_i + 1$ for $i =1,2$.  Let $x_1$ and $x_2$ be probability vectors indexed by the elements of $\tilde{C}_1$ and $\tilde{C}_2$, respectively.  For each such pair $(S_1,S_2)$ we use the fact that linear programs are polynomial-time solvable to find $x_1$ and $x_2$ such that the following linear constraints are satisfied, or prove that no such vectors exist \cite{bt:ilo}.
\begin{equation}
\label{eq:nashlp}
\begin{split}
\left[x_1 U_2\right]_{s_2} \geq \left[x_1 U_2\right]_{t_2} & \text{ for all }s_2\in S_2, t_2\in\tilde{C}_2 \\
\left[U_1 x_2\right]_{s_1} \geq \left[U_1 x_2\right]_{t_1} & \text{ for all }s_1\in S_1, t_1\in\tilde{C}_1 \\
x_i(s_i) \geq 0 & \text{ for all } s_i\in S_i, i = 1,2 \\
x_i(s_i) = 0 & \text{ for all } s_i\in \tilde{C}_i\setminus S_i, i = 1,2 \\
\sum_{s_i\in S_i} x_i(s_i) = 1 & \text{ for } i = 1,2
\end{split}
\end{equation}
(There are many redundant constraints here which could easily be removed, but we have presented the constraints in this form for simplicity.)  Any feasible point for any pair $(S_1,S_2)$ is a Nash equilibrium of the sampled game and an $\epsilon$-equilibrium of the original game.  The algorithm will find at least one such point.
\end{algorithm}

\begin{proof}
For the purpose of analyzing the complexity of the algorithm we will view the Lipschitz constants as fixed, even as $\rho$ varies.  Let $\tilde{u}_i$ be the payoffs of the sampled game and suppose $\sigma$ is a Nash equilibrium of the sampled game.  Choose any $s_i\in C_i$ and let $s'_i$ be an element of $\tilde{C}_i$ closest to $s_i$, so $\lvert s_i - s'_i\rvert \leq \frac{\epsilon}{L_i}$.  Then
\begin{equation*}
\begin{split}
    u_i(s_i,& \sigma_{-i}) - u_i(\sigma) \\
    & \leq u_i(s_i,\sigma_{-i}) - u_i(s'_i,\sigma_{-i}) + u_i(s'_i,\sigma_{-i}) - u_i(\sigma) \\    & \leq \lvert u_i(s_i,\sigma_{-i}) - u_i(s'_i,\sigma_{-i}) \rvert + \tilde{u}_i(s'_i,\sigma_{-i}) - \tilde{u}_i(\sigma) \\
    & \leq L_i\frac{\epsilon}{L_i} + 0 = \epsilon
\end{split}
\end{equation*}
so $\sigma$ is automatically an $\epsilon$-equilibrium of the original separable game.  Thus it will suffice to compute a Nash equilibrium of the finite sampled game.

To do so, first compute or bound the rank $\rho$ of the original
separable game using Theorem \ref{thm:rankcomp} or
\ref{thm:rankbound}.  By Theorem \ref{thm:seprank} and Corollary
\ref{cor:ranksamp}, the sampled game has a Nash equilibrium in which
player $i$ mixes among at most $\rho_i + 1$ pure strategies,
independent of how large $|\tilde{C}_i|$ is.  The separability
assumption is fundamental because without it we would not obtain
this uniform bound independent of $|\tilde{C}_i|$.  The number of
possible choices of at most $\rho_i + 1$ pure strategies from
$\tilde{C}_i$ is
\begin{equation*}
\sum_{k=1}^{\rho_i + 1} \binom{|\tilde{C}_i|}{k} \leq \binom{|\tilde{C}_i|+\rho_i}{1+\rho_i} = \binom{|\tilde{C}_i| + \rho_i}{|\tilde{C}_i| - 1},
\end{equation*}
which is a polynomial in $|\tilde{C}_i|\propto \frac{1}{\epsilon}$
for fixed $\rho$ and a polynomial in the components of $\rho$ for
fixed $\epsilon$.  This leaves the step of checking whether there
exists an equilibrium $\sigma$ for a given choice of $S_i =
\supp(\sigma_i)\subseteq\tilde{C}_i$ with $|S_i|\leq \rho_i + 1$ for
each $i$, and if so, computing such an equilibrium.  Since the game
has two players, the set of such equilibria for given supports is
described by a number of linear equations and inequalities which is
polynomial in $\frac{1}{\epsilon}$ for fixed $\rho$ and polynomial
in the components of $\rho$ for fixed $\epsilon$; these equations
and inequalities are given by \eqref{eq:nashlp}.  Since linear programs are polynomial-time solvable, we can find a feasible solution to
such inequalities or prove infeasibility in polynomial time.  The
two player assumption is key at this step, because with more players
the constraints would fail to be linear or convex and we could no
longer use a polynomial time linear programming algorithm.

Thus we can check all supports and find an $\epsilon$-equilibrium of the sampled game in polynomial time as claimed.
\end{proof}

We will now consider weakening Assumption
\ref{assumption:simplifying}.  The Lipschitz condition could be
weakened to a H\"{o}lder condition and the same proof would work,
but it seems that we must require some quantitative bound on the
speed of variation of the utilities in order to bound the running time
of the algorithm.  Also, the strategy space could be changed to any
compact set which can be efficiently sampled, e.g.\ a box in $\Rm^n$.
However, for the purpose of asymptotic analysis of the algorithm,
the proof here only goes through when the Lipschitz constants and
strategy space are fixed.  A more complex analysis would be required
if the strategy space were allowed to vary with $\rho$, for example.

It should be noted that the requirement that the
strategy space be fixed for asymptotic analysis means that Theorem
\ref{thm:epseq} does not apply to finite games, at least not if the
number of strategies is allowed to vary.  For the sake of comparison
and completeness we state the best known $\epsilon$-equilibrium
algorithm for finite games below.

\begin{theorem}[Lipton et al.\ \cite{lmm:plgss}]
\label{thm:lmmalg}
There exists an algorithm to compute an $\epsilon$-equilibrium of an $m$-player finite game with $n$ strategies per player which is polynomial in $\frac{1}{\epsilon}$ for fixed $m$ and $n$, polynomial in $m$ for fixed $n$ and $\epsilon$, and quasipolynomial in $n$ for fixed $\epsilon$ and $m$ (assuming the payoffs of the games are uniformly bounded).
\end{theorem}

In the case of two-player separable games which we have considered,
the complexity of the payoffs is captured by $\rho$, which is bounded by the cardinality of the strategy spaces in two-player finite games.  Therefore in finite games the complexity of the payoffs and the complexity of the strategy spaces are intertwined, whereas in games with infinite strategy spaces they are decoupled.  The best known algorithm for finite games stated in Theorem \ref{thm:lmmalg} has quasipolynomial dependence on the complexity of the game.   Our algorithm is interesting because it has polynomial dependence on the complexity of the payoffs when the strategy spaces are held fixed.  In finite games this type of asymptotic analysis is not possible due to the coupling between the two notions of complexity of a game, so a direct comparison between Theorem \ref{thm:epseq} and Theorem \ref{thm:lmmalg} cannot be made.  

\section{Conclusions}

We have shown that separable games provide a natural setting for the
study of games with payoffs satisfying a low-rank condition.  This
level of abstraction allows the low-rank results of Lipton et al.
\cite{lmm:plgss} to be extended to infinite strategy spaces and
multiple players. Since the rank of a separable game gives a bound
on the cardinality of the supports of equilibria for any sampled
version of that separable game, approximate equilibria can be
computed in time polynomial in $\frac{1}{\epsilon}$ by discretizing
the strategy spaces and applying standard computational techniques
for low-rank games.

Other types of low-rank conditions have been studied for finite
games, for example Kannan and Theobald have considered the condition
that the sum of the payoff matrices be low-rank \cite{kt:gfr}.  It
is likely that that the discretization techniques used here can be
applied in an analogous way to yield results about computing
approximate equilibria of continuous games when the sum of the
payoffs of the players is a separable function.

There also exist many computational techniques for finite games
which do not make low-rank assumptions.  It may be possible to
extend some of these techniques directly to separable games to yield
algorithms for computing exact equilibria of separable games.  Such
an extension would likely require an explicit description of the
moment spaces in terms of inequalities rather than the description
given above as the convex hull of the set of moments due to pure
strategies.  In the case of two-player polynomial games, such an
explicit description is known to be possible using linear matrix
inequalities and has been applied to zero-sum polynomial games by
Parrilo \cite{pp:polygames}.  While the lack of polyhedral structure
in the moment spaces would most likely prohibit the use of a
Lemke-Howson type algorithm, a variety of other finite game
algorithms may be extendable to this setting; see McKelvey and
McLennan for a survey of such algorithms \cite{mm:ce}.

Finally, there exist a variety of other solution concepts for strategic form games which may be amenable to analysis and computation in the case of separable games, and in particular in polynomial games.  Preliminary results on computation of correlated equilibria appear in \cite{s:mastersthesis, spo:cdc07}.  For a correlated equilibrium version of the rank bounds on Nash equilibria of separable games presented above, see \cite{s:mastersthesis}.  We leave the extension to other solution concepts, in particular iterated elimination of strictly dominated strategies, for future work.

\newpage

\bibliographystyle{plain}
\bibliography{../../references}
\end{document}